\documentclass[11pt]{article}
\usepackage{amssymb}
\usepackage{amsmath} 

   \csname @addtoreset\endcsname{equation}{section}
    \csname @addtoreset\endcsname{figure}{section}
    \csname @addtoreset\endcsname{table}{section}

\newcounter{thanksnum}
\def\thanksnumber#1
{\setcounter{thanksnum}{\value{footnote}}\setcounter{footnote}{#1}%
                     \addtocounter{footnote}{-1}\footnotemark
                     \setcounter{footnote}{\value{thanksnum}}}
\def\newtheoremz#1{\@ifnextchar[{\@othmz{#1}}{\@nthmz{#1}}}

\def\@nthmz#1#2{%
\@ifnextchar[{\@xnthmz{#1}{#2}}{\@ynthmz{#1}{#2}}}

\def\@xnthmz#1#2[#3]{\expandafter\@ifdefinable\csname #1\endcsname
{\@definecounter{#1}\@addtoreset{#1}{#3}%
\expandafter\xdef\csname the#1\endcsname{\expandafter\noexpand
  \csname the#3\endcsname \@thmcountersepz \@thmcounterz{#1}}%
\global\@namedef{#1}{\@thmz{#1}{#2}}\global\@namedef{end#1}{\@endtheoremz}}}

\def\@ynthmz#1#2{\expandafter\@ifdefinable\csname #1\endcsname
{\@definecounter{#1}%
\expandafter\xdef\csname the#1\endcsname{\@thmcounterz{#1}}%
\global\@namedef{#1}{\@thm{#1}{#2}}\global\@namedef{end#1}{\@endtheoremz}}}

\def\@othmz#1[#2]#3{\expandafter\@ifdefinable\csname #1\endcsname
  {\global\@namedef{the#1}{\@nameuse{the#2}}%
\global\@namedef{#1}{\@thmz{#2}{#3}}%
\global\@namedef{end#1}{\@endtheoremz}}}

\def\@thmz#1#2{\refstepcounter
    {#1}\@ifnextchar[{\@ythmz{#1}{#2}}{\@xthmz{#1}{#2}}}

\def\@xthmz#1#2{\@begintheoremz{#2}{\csname the#1\endcsname}\ignorespaces}
\def\@ythmz#1#2[#3]{\@opargbegintheoremz{#2}{\csname
       the#1\endcsname}{#3}\ignorespaces}

\def\@thmcounterz#1{\noexpand\arabic{#1}}
\def\@thmcountersepz{.}
\def\@begintheoremz#1#2{ \trivlist \item[\hskip \labelsep{\bf #1\ #2}]}
\def\@opargbegintheoremz#1#2#3{ \trivlist
      \item[\hskip \labelsep{\bf #1\ #2\ (#3)}]}
\def\@endtheoremz{\endtrivlist}

\newtheorem{theorem}{Theorem}[section]

\newtheorem{corollary}{Corollary}[section]

\newtheorem{remark}{Remark}[section]

\newtheorem{example}{Example}[section]

\def\e{\varepsilon}

\def\defi{\stackrel{{\scriptscriptstyle \Delta}}{=}}

\def\d{\delta}
\def\o{\omega}
\def\O{\Omega}

\def\esssup{\mathop{\rm ess\, sup}}

\def\R{{\bf R}}
\def\E{{\bf E}}
\def\P{{\bf P}}

\def\\x{{\bf \x}}

\def\s{\delta}

\def\ww{\widetilde}
\def\t{\theta}
\def\oo{\bar}
\def\s{\sigma}

\def\p{\partial}

\def\B{\mathcal{B}}

\def\PP{\mathcal{P}}

\def\Z{\mathcal{Z}}
\def\PP{\mathcal{PP}}
\def\X{\mathcal{X}}

\def\U{\mathcal{U}}
\def\V{\mathcal{V}}

\def\F{\mathcal{F}}

\newcommand{\be}{\begin{equation}}
\newcommand{\ee}{\end{equation}}
\newcommand{\bd}{\begin{displaymath}}
\newcommand{\ed}{\end{displaymath}}
\newcommand{\ba}{\begin{array}{ll}}
\newcommand{\ea}{\end{array}}
\newcommand{\baa}{\begin{eqnarray}}
\newcommand{\eaa}{\end{eqnarray}}
\newcommand{\baaa}{\begin{eqnarray*}}
\newcommand{\eaaa}{\end{eqnarray*}}
\font\sm=cmr10

\def\B\x{{\scriptscriptstyle B\x}}

\def\oo{\bar}

\def\y{\mu}

\def\logs{{\scriptscriptstyle log}}

\def\QQ{{ Q}}

\def\U{{\cal U}}

\def\QQ{\\PP_\t }
\def\X{{\cal X}}
\def\Y{{\cal Y}}

\def\QQ{{\oo x}}
\def\PP{{\rm x}}

\def\x{x}
\date{Submitted: November 4, 2013. Revised:  February 6, 2015} 
\title{
On strong binomial approximation for stochastic processes with applications for financial modelling 
}
\author{
Nikolai Dokuchaev \index{Corresponding address: Department of
Mathematics \& Statistics, Curtin University, GPO Box U1987, Perth,
6845 Western
Australia,  email N.Dokuchaev@curtin.edu.au, tel. 61 8 92663144.} \\
{\sm Department of
Mathematics \& Statistics, Curtin University,}\vspace{-0.5cm} \\
\vspace{-0.5cm} {\sm  GPO Box U1987, Perth, 6845 Western Australia}}
\begin{document}
\maketitle
\begin{abstract}
This paper considers binomial approximation of continuous time
stochastic processes. It is  shown that, under some mild
integrability conditions, a process can be approximated in mean
square sense and in other strong metrics by binomial processes,
i.e., by processes with fixed size binary increments at sampling
points. Moreover, this approximation can be causal, i.e., at every
time it requires  only  past historical values of the underlying
process. In addition, possibility of approximation of solutions of
stochastic differential equations by solutions of ordinary equations
with binary noise is established. Some consequences for the
financial modelling and options pricing models are discussed.
\par
\index{{\bf JEL classification}:
B23, 
C02, 
C51, 
G13 
}
\end{abstract}
\section{Introduction}
This paper considers approximation of continuous time stochastic
processes by binomial  piecewise affine processes. Usually, these
problems are studied in the framework of the Functional Limit
Theorems and weak convergence, i.e., convergence in distributions.
 The
classical result is the Donsker Theorem that establishes  weak
convergence  for some particular processes with given distributions
(Donsker \cite{Don}, 1952). Currently, there are many results on the
weak convergence for many types  of underlying continuous time
processes and approximating discrete time processes; see, e.g.,
\cite{Bil,Bor,Ded,I62,Ni,Tu,vV}, \index{Billingsley (1968),
Borovkova {\em et al} (2001), Dedecker and Prieur (2005), Ibragimov
(1962), Nickl {\em et al} (2013), Tudor and Torres (2007), van der
Vaart and van Zanten (2005),} and the bibliography there. As far as
we know, the convergence of discrete time processes to continuous
time processes in the strong sense  was not considered in the
literature, including approximation in the mean square sense, in
$L_q$-norm, or in probability.

The possibility to approximate a continuous time process by discrete
time processes  appears to be important for applications in
financial modelling.  This possibility  allows to replace the
options pricing for continuous time market models by the options
pricing for discrete time market models. The models  based on
binomial processes, i.e., with fixed size increments, are especially
important for this purpose. The reason for this is that the
corresponding discrete time market models are usually complete and
allow  uniquely defined prices for the derivatives. These prices can
be conveniently calculated by the so-called binomial trees method
which represents a special case of the finite difference methods for
PDEs. There are many works on this topic; see, e.g.,
\cite{AD,Amin,He,Ne}, \index{ Akyildirim {\em et al}  (2012), Amin
(1991),
 Heath {\em et al}  (1990),
Nelson and Ramaswamy (1990),} and the bibliography there. Again, the
approximation for the financial models was considered in the weak
sense (i.e., in distributions).

With respect to  the options pricing problem, the particular
distributions of the approximating binomial processes is not really
important, since the pricing formula is based on an artificial
martingale (or risk neutral)  measure rather than on the  measure
generated by the observed prices. The only part of the Donsker
Theorem used in this framework was the existence of the binomial
approximating processes. In the present paper, we address only this
aspect of the Donsker Theorem: the existence of binomial
approximations, without specifications of their distributions.

We consider  binomial approximation of continuous time stochastic
processes  in $L_q$-norm, where $q\in[1,+\infty)$; this is a strong
convergence  that includes convergence in mean square and implies
convergence in probability.   In Section \ref{secD}, we show that a
general $L_q$-integrable stochastic process can be approximated in
$L_q$-norm by  pathwise continuous processes with fixed size binary
increments for the sampling points; in particular, continuous It\^o
processes and processes with jumps are covered (Theorem \ref{ThM}). Moreover, we show
that this approximation can be {\em causal}, i.e., the value of an
approximating process  at each time is calculated using only the
past historical values of the underlying process; in other words,
the approximating process is adapted to the filtration generated by
the underlying process.
 This can be interpreted as
a strong version of Donsker Theorem.

It can be noted that we consider approximation in a different setting
than in the cited papers on weak convergence of discrete time processes, where some particular distributions  were assumed for the  underlying continuous
time process and for the approximating processes. We do not assume a particular distribution or certain dynamic properties such as independence or correlation of the increments.
We rather suggest an algorithm that allows to
construct the approximating processes of the prescribed binomial type
from the current observations of the underlying process. Therefore, our
approximation  result is not in the framework of the Functional Limit Theorem;  it does not establish convergence of particular distributions.

In a more general setting, we consider approximation of a stochastic
process by solutions of ordinary differential equations with a given
drift coefficient and with binary noise (Theorems \ref{Th2} and
\ref{Th4}). In particular, we found that the solution of a
stochastic It\^o equation can be approximated by solutions of
ordinary differential equations with the same  drift coefficient and
with binary noise replacing  the driving Wiener process (Corollary
\ref{corr1}). Currently, there are many methods of discretization of
stochastic differential equations such as  Euler-Maruyama
discretization; see,  e.g., \cite{AK,Hi,K}. \index{Abramov {\em et
al} (20111), Higham {\em et al} (2002) or Kloeden and Platen
(1992).} Theorems \ref{Th2} and  \ref{Th4} could be a useful
addition to the existing methods of discretization of stochastic
differential equations.

It appears that these approximation results  have some implications for financial modelling and for the general pricing
theory. To illustrate this, we show  that existing of binomial approximations with fixed rate of changes
implies that, for a  incomplete continuous time market
model, there exists a  complete  model
    such that these two models are statistically indistinguishable,
    given the presence of any non-zero errors in the measurements.
    This feature is non-trivial. It is  well known that the market completeness is
not a robust property: there are  arbitrarily small random contaminations  of the coefficients that
can convert a complete market model into a incomplete one. We are presenting an "inverse" property:  the market incompleteness is
also non-robust, meaning that there exist arbitrarily small contaminations that can convert an incomplete model
into a complete one.

The paper is organized as follows. In Section \ref{secD}, we
consider approximation by continuous binomial processes  and by the
solutions of related ODEs with binary noise inputs. In Section
\ref{secM}, we consider some useful modifications of the main
result, including  approximation  by piecewise constant approximating processes and by the
solutions of related ODEs with binary jumps. In addition, we discuss
in Section \ref{secM}  approximation of positively valued processes
by   the processes similar to the prices in Cox-Ross-Rubinstein
model, and a possibility of dynamic replication of the changes of
the diffusion coefficient. In Section \ref{secF},  we discuss some
implications for financial modelling and pricing theory.
\section{The problem setting and the main result}\label{secD}
Consider a
standard probability space $(\O,\F,\P)$, where $\O$ is a set of
elementary events,
$\P$ is a probability measure, and  $\F$ is a $\P$-complete $\s$-algebra of events.

Let $T>0$ be given, $q\in[1,+\infty)$. Let $X$ be the set of   real
valued stochastic processes such that $\|x\|_{X}\defi\left(\E\int_0^T|x(t)|^qdt\right)^{1/q}<+\infty$
for $x\in X$.

Let
$X_c$ be the set of all processes $x\in X$ such that for any $x\in X_c$ there exists $\t=\t(x(\cdot))\in[0,T)$
such that the mapping $x:[\t,T]\to L_q(\O,\F,\P)$ is continuous.

For $x\in X_c$, we denote
$\|x\|_{X_c}\defi\|x\|_X+\left(\E|x(T)|^q\right)^{1/q}$. Clearly, this value is uniquely defined for
any $x\in X_c$.

Let $ x \in X$ be given, and let $\F_t$ be the filtration generated
by $ x (t)$.

Let $\Y_n=\Y_n(x(\cdot))$ be the set of pathwise continuous $\F_t$-adapted real valued processes
$y(t)$ such
that $y(0)= x (0)$ and that there exists $d>0$ such that  either $y(t)=y(t_k)+d(t-t_k)$
for $t\in[t_k,t_{k+1})$
or $y(t)=y(t_k)-d(t-t_k)$ for $t\in[t_k,t_{k+1})$, where $t_k=kT/n$,
$k=0,...,n$, $n=1,2,...$.
The sequence $\{y(t_k)\}$ represents a path of a so-called binomial tree.

Let $\Y=\cup_{n\ge 1}\Y_n$.

Our main result can be formulated as the following.
\begin{theorem}\label{ThM}
\begin{enumerate}
\item
For any $ x \in X$ and $\e>0$, there exists $y\in\Y$ such that \baaa
\| x -y\|_{X}\le \e. \eaaa
\item
 For any $ x \in X_c$ and $\e>0$, there exists $y\in\Y$ such that \baaa
\| x -y\|_{X_c}\le \e. \eaaa
\end{enumerate}
\end{theorem}
\par
{\em Proof.} Without a loss of generality, we assume that $ x (t)$ is
defined for $t<0$ and that $ x (t)= x (0)$ for $t<0$.

Let \baaa \QQ_m(t)\defi \min(\max( x (t),-m),m),\quad  m=1,2,.... \eaaa
Clearly, $|\QQ_m|\le m$ and $\QQ_m(t)= x (t)$ if and only if $| x (t)|\le m$.

Let \baaa \PP_{m,p}(t)\defi
\frac{1}{\e_p}\int_{t-\e_n}^t\QQ_m(s)ds,\quad \e_p=1/p,\quad p=1,2,...
\eaaa Clearly, this process is pathwise absolutely continuous and such that
$$
\esssup_t|d\PP_{m,p}(t)/dt|\le 2\e_k^{-1}\sup_{t\in[0,T]}|\QQ_m(t)|\le 2mp.$$

Let any $K\ge 0$ be selected.  Let $M_{m,p}\defi 2mp+K.$ (For the
proof of Theorem \ref{ThM}, it suffices to use $K=0$; we need
$K>0$ for the proof of the next theorem).

Let us consider $n=1,2,...$. Let $t_k=kT/n$, $k=0,...,n$. Let the process
$y(t)=y_{n,m,p}(t)$ be defined such that $y(0)=\PP_{m,p}(0)$,
$y(t)=y(t_k)+M_{m,p}(t-t_k)$ for $t\in[t_k,t_{k+1})$ if $y(t_k)\le
\PP_{m,p}(t_k)$, and $y(t)=y(t_k)-M_{m,p}(t-t_k)$ for
$t\in[t_k,t_{k+1})$, if $y(t_k)\ge \PP_{m,p}(t_k)$. Clearly, $y\in \Y_p$.

Let $\d=\d(n)=t_{k+1}-t_k=T/n$. Let us show that \baa
|y(t)-\PP_{m,p}(t)|\le 2M_{m,p}\d, \quad t\in[0,T]. \label{ineq}\eaa

 Clearly, (\ref{ineq}) holds for $t=t_0$.
 It suffices to  show that if $|y(t_k)-\PP_{m,p}(t_k)|\le 2M_{m,p}\d$ then \baa
|y(t)-\PP_{m,p}(t)|\le 2M_{m,p}\d, \quad  t\in[t_k,t_{k+1}], \quad
k=0,...,n. \label{ineq0}\eaa

Let $M_1\defi  2M_{m,p}\d$.  For $t\in[t_k,t_{k+1}]$, let
$M=M_{m,p}(t-t_k)$. We have to consider several possible scenarios.

\def\pp{\PP_{m,p}(t)}
\def\y1{y(t)}
\begin{enumerate}
\item
 Assume
that $\PP_{m,p}(t_k) -y(t_k)\in[0,M_1]$ and
 $\pp\ge \PP_{m,p}(t_k)$. In
this case, $\pp=\PP_{m,p}(t_k)+M-\e$, where $\e\in[0,M]$. Hence
\baaa\pp -\y1&=&\PP_{m,p}(t_k)+M-\e-y(t_k)-M\\&=&\PP_{m,p}(t_k)-y(t_k)-\e\in[-\e,M_1-\e].\eaaa
\item
Assume that $\PP_{m,p}(t_k)-y(t_k)\in(M_1,2M_1]$ and $\pp \ge
\PP_{m,p}(t_k)$. In this case, $\pp =\PP_{m,p}(t_k)+M-\e$ again, where
$\e\in[0,M]$. Hence
\baaa\pp -\y1&=&\PP_{m,p}(t_k)+M-\e-y(t_k)-M\\&=&=\PP_{m,p}(t_k)-y(t_k)-\e\in(M_1-\e,2M_1-\e].\eaaa
\item
Assume that $\PP_{m,p}(t_k) -y(t_k)\in[0,M_1]$ and
 $\pp < \PP_{m,p}(t_k)$. In
this case, $\pp =\PP_{m,p}(t_k)-M+\e$, where $\e\in(0,M]$. Hence
\baaa\pp -\y1&=&\PP_{m,p}(t_k)-M+\e-y(t_k)-M\\&=&\PP_{m,p}(t_k)-y(t_k)-2M+\e\in[-2M+\e,M_1-2M+\e].\eaaa

\item Assume that $\PP_{m,p}(t_k)-y(t_k)\in(M_1,2M_1]$ and $\pp <
\PP_{m,p}(t_k)$. In this case, $\pp =\PP_{m,p}(t_k)-M+\e$ again, where
$\e\in(0,M]$. Hence
\baaa
\pp -\y1&=&\PP_{m,p}(t_k)-M+\e-y(t_k)-M\\&=&\PP_{m,p}(t_k)-y(t_k)-2M+\e\in(M_1-2M+\e,2M_1-2M+\e].\eaaa

\item Assume that $\PP_{m,p}(t_k)-y(t_k)\in[-M_1,0)$ and $\pp \le
\PP_{m,p}(t_k)$. In this case, $\pp =\PP_{m,p}(t_k)-M+\e$, where
$\e\in[0,M]$. Hence
\baaa
\pp -\y1&=&\PP_{m,p}(t_k)-M+\e-y(t_k)+M\\&=&\PP_{m,p}(t_k)-y(t_k)+\e\in[-M_1+\e,\e).\eaaa
\item Assume that $\PP_{m,p}(t_k)-y(t_k)\in[-2M_1,-M_1)$ and $\pp \le
\PP_{m,p}(t_k)$. In this case, $\pp =\PP_{m,p}(t_k)-M+\e$, where
$\e\in[0,M]$. Hence
\baaa\pp -\y1&=&\PP_{m,p}(t_k)-M+\e-y(t_k)+M\\&=&\PP_{m,p}(t_k)-y(t_k)+\e\in[-2M_1+\e,-M_1+\e).\eaaa

\item Assume that $\PP_{m,p}(t_k)-y(t_k)\in[-M_1,0)$ and $\pp >
\PP_{m,p}(t_k)$. In this case, $\pp =\PP_{m,p}(t_k)+M-\e$, where
$\e\in[0,M)$. Hence
\baaa\pp -\y1&=&\PP_{m,p}(t_k)+M-\e-y(t_k)+M\\&=&\PP_{m,p}(t_k)-y(t_k)+2M-\e\in[2M-M_1-\e,2M-\e).\eaaa

\item Assume that $\PP_{m,p}(t_k)-y(t_k)\in[-2M_1,-M_1)$ and $\pp  >
\PP_{m,p}(t_k)$. In this case, $\pp =\PP_{m,p}(t_k)+M-\e$, where
$\e\in[0,M)$. Hence
\baaa\pp -\y1&=&\PP_{m,p}(t_k)+M-\e-y(t_k)+M\\&=&\PP_{m,p}(t_k)-y(t_k)+2M-\e\in[2M-2M_1-\e,2M-M_1+\e).\eaaa
\end{enumerate}

By the assumptions, $M_1\le M$. It follows from (i)-(viii) that
$\pp -\y1\in(-2M_1,2M_1)$ for all possible scenarios. Hence
(\ref{ineq0}) holds and therefore (\ref{ineq}) holds.

We are now in the position to complete the proof.

 Let
$\|\cdot\|_{\X}\defi \|\cdot\|_X$ for the proof of statement  (i), and
let $\|\cdot\|_{\X}\defi \|\cdot\|_{X_c}$ for the proof of statement (ii).

By the Lebesgue's Dominated Convergence Theorem, we have that \baaa
\| x -\QQ_m\|_{\X}\to 0\quad \hbox{as}\quad m\to +\infty. \eaaa By
the Lebesgue's Dominated Convergence Theorem again, for any $m$,
\baaa \|\QQ_m-\PP_{m,p}\|_{\X}\to 0\quad \hbox{as}\quad p\to
+\infty. \eaaa In addition, it follows from (\ref{ineq}) that, for
any $m$ and $p$, \baaa \|\PP_{m,p}-y_{n,m,p}\|_{\X}\to 0\quad
\hbox{as}\quad n\to +\infty. \eaaa
\par
 Let $\e>0$ be
given. It suffices to show that there  exists $n,m,p$ such that
\baa\|x-y\|_{\X}\le\e,\label{proof}\eaa  for $y=y_{n,m,p}$
constructed  as described above.

Let $m$ be such that \baa \| x -\QQ_m\|_{\X}\le\frac{\e}{3}.
\label{mA}\eaa Further, let $p=p(m)$ be such that \baa
\|\QQ_m-\PP_{m,p}\|_{\X}\le\frac{\e}{3}. \label{nA}\eaa Finally, let
$n=n(m,p)$ be such that $2T^{1/q}M_{m,p}\d\le\e/3$, where
$\d=\d(n)=T/n$. In this case, it follows from (\ref{ineq}) that
 \baa
\|\PP_{m,p}-y_{n,m,p}\|_{\X}\le\frac{\e}{3}. \label{pA}\eaa
Estimates (\ref{mA})-(\ref{pA}) imply (\ref{proof}). This completes
the proof. $\Box$
\begin{remark} Theorem \ref{ThM} does not require any information on the evolution and the distribution of
$ x (\cdot)$. Respectively, this  theorem does not suggest how to
select the set $(n,m,p)$ for a given $\e$. If  $(n,m,p)$ is
selected, then the values of the process $y(t)$ at any time $t$ are
computed using the historical observations of $ x (s)|_{s\le t}$,
according to the algorithm described in the proof of Theorem
\ref{ThM}; in other words, the approximating process is
$\F_t$-adapted, and its choice is causal.
\end{remark}
\begin{remark}  If $ x (t)$ is a bounded process, then one can use $ x (t)$ directly instead of
$\QQ_m(t)$. If $\QQ_m(t)$ is absolutely continuous and
$\oo c=\esssup_t|d\QQ_m(t)/dt|<+\infty$, then one can construct $y(t)$  using
$\QQ_m(t)$ instead of $\PP_{m,p}(t)$ and $\oo c$ instead of $M_{m,p}$.
\end{remark}
We remind that the paths of the processes  $y\in \Y_n$ are piecewise affine and continuous; therefore, these paths  are  absolutely continuous, left-differentiable and right-differentiable; they are differentiable at all $t\neq t_k$, $k=0,...,n$, $t_k=kT/n$.
\par
We denote by $D^{\pm}_t$ the left-hand side derivative or right-hand side derivative respectively.
\begin{example}\label{exa1} {\rm Theorem \ref{ThM} is oriented on non-differentiable stochastic
processes. However, it will be useful to expose some properties of
the processes $y(t)$ using the following toy examples.
\begin{enumerate} \item
 Assume that $ x (t)\equiv 0$. In this case, the approximating processes
 $y(t)$ defined in Theorem \ref{ThM} can be considered with fixed $m=1$; they are
periodic functions oscillating about $ x (t)$ and such that
$|D^{\pm}_t y(t)|\equiv c$, where $c>0$. This $c$ can be selected arbitrarily.
\item
Assume that $ x (t)=0$ for $t<T/2$ and  $ x (t)=1$ for $t\ge T/2$. In this case, it suffices to use fixed $m=1$ again. The approximating processes
 $y(t)$ defined in Theorem \ref{ThM} oscillate about
zero for $t<T/2$, and
 $|D^{\pm}_t y(t)|\to+\infty$ as $\|y- x \|_{X}\to 0$, i.e., as $p\to
+\infty$ and $n\to+\infty$. This is because
$\esssup_{t\in[0,T]}|d\PP_{m,p}(t)/dt|\to +\infty$ as $p\to +\infty$.
 Since $|D^{\pm}_t y(t)|$ is constant in $t\in[0,T]$, this shows that the approximation suggested does not track the rate of change for
the underlying process.
\end{enumerate}}
\end{example}
\subsection*{Approximation via solutions of ODEs with binary noise}
\label{secO}

Let $f(x,t):\R\times[0,T]\to\R$ be a continuous function
such that $|f(x,t)|+|\p f(x,t)/\p x|\le c_f$ for some $c_f>0$.

\par
Again, we assume that $ x \in X$ is given, and that $\F_t$ is the filtration generated by $ x $.

\par

Let $\U_n$ be the set of real
valued processes $u(t)$ such that $u(0)= x (0)$ and that \baa
u(t)=u(0)+\int_0^tf(u(s),s)ds+y(t), \label{u}\eaa where $y\in\Y_n$.

Clearly, $u$ is uniquely defined  pathwise continuous and
$\F_t$-adapted process that satisfy the ordinary differential
equation (ODE) \baaa &&\frac{du}{dt}(t)=f(u(t),t)+\eta(t)
\label{u2}\eaaa with binary noise $\eta(t)=D_t^{\pm}y(t)$.
The choice of the right hand  derivative $D_t^+$ or left-hand derivative $D_t^-$
here does not affect the solution of the ODE, since these derivatives coincides
everywhere except the points $t_k=\d k$, $\d=T/n$.

The process  $\eta$  is adapted to the
filtration generated by $ x $, and its properties are defined by the
properties of $ x $.

Let $\U=\cup_{n\ge 1}\U_n$.

\begin{theorem}\label{Th2}
\begin{enumerate}
\item
For any $ x \in X$ and $\e>0$, there exists $u\in\U$ such that \baaa
\| x -u\|_{X}\le \e. \eaaa
\item
 For any $ x \in X_c$ and $\e>0$, there exists $u\in\U$ such that \baaa
\| x -u\|_{X_c}\le \e. \eaaa
\end{enumerate}
\end{theorem}
\par
Clearly, Theorem \ref{Th2} applied with $f\equiv 0$  gives
Theorem \ref{ThM}; therefore, Theorem \ref{Th2} represents a  generalization of Theorem
\ref{ThM}.
\par
{\em Proof of Theorem \ref{Th2}.} We use the processes $\QQ_m$ and
$\PP_{m,p}$ from the proof of Theorem \ref{ThM}. We modify the
construction of $y(t)=y_{n,m,p}(t)$ as the following. We select
$K=\sup_{x,t}|f(x,t)|$, i.e., $M_{m,p}=2mp+\sup_{x,t}|f(x,t)|$.  For
 $n\in\{1,2,...\}$, we set $t_k=kT/n$, $k=0,...,n$,  and define step
functions $\t(s)$ such that $\t(s)=t_k$ if $s\in[t_k,t_{k+1})$.

Let us construct the processes $ r (t)= r _{n,m,p}(t)$, $y(t)=y_{n,m,p}(t)$, and
$u(t)=u_{n,m,p}(t)$ as the following. We assume that  $ r (0)=y(0)=0$,  $u(0)= x (0)$, and \baaa && r (t)\defi \PP_{m,p}(t)- x (0)-\int_0^tf(u(\t(s)),s)ds, \\
&& u(t)\defi u(0)+\int_0^tf(u(s),s)ds+y(t).\eaaa
 Here the process $y(t)$ is
defined as the following:  $y(t)=y(t_k)+M_{m,p}(t-t_k)$ for
$t\in[t_k,t_{k+1})$ if $y(t_k)\le  r (t_k)$, and
$y(t)=y(t_k)-M_{m,p}(t-t_k)$ for $t\in[t_k,t_{k+1})$, if $y(t_k)>
 r (t_k)$.

Clearly,
$y\in \Y_n$, and the processes $ r $, $u$, and $y$, can be constructed
consequently on the intervals $[t_k,t_{k+1}]$, $k=0,1,2,...$.

Let $\d=t_{k+1}-t_k=T/n$. Similarly to (\ref{ineq}), we obtain that
\baaa |y(t)- r (t)|\le 2M_{m,p}\d, \quad t\in[0,T]. \label{ineqU}\eaaa
Let
 \baaa
\ww  r (t)=\ww  r _{n,m,p}(t)\defi  x (t)- x (0)-\int_0^tf(u(s),s)ds. \eaaa
Clearly, \baaa |\ww  r (t)- r (t)|&\le& \int_0^T|f(u(\t(s)),s)-
f(u(s),s)|ds\\&\le& T\sup_{s}|f(u(\t(s)),s)- f(u(s),s)|
\\&\le& c_fT\sup_s|u(\t(s))- u(s)|\\&\le&
c_fT\sup_s\left|\int_{\t(s)}^s f(u(r),r)dr+ y(s)-y(\t(s))\right|
\\ &\le &c_f(c_f\d+M_{m,p}\d)=c_f(c_f+M_{m,p})\frac{T}{n}.\eaaa

It follows that, for all $t$, \baaa |\PP_{m,p}(t)-u(t)|=|\ww   r (t)-y(t)|\le |\ww
 r(t)- r (t)|+| r (t)-y(t)|\to 0 \quad\hbox{as}\quad n \to +\infty. \eaaa
The remaining part of the proof follows the proof of
Theorem \ref{ThM}. $\Box$
\begin{corollary}\label{corr1} Let $f(x,t):\R\times[0,T]\to\R$ and  $b(x,t):\R\times[0,T]\to\R$ be continuous bounded functions such that the derivatives $\p f(x,t)/\p x$ and  $\p b(x,t)/\p x$ are also bounded.  Let $w(t)$ be
a standard Wiener process, and let the evolution of $ x $ be described by
the It\^o equation  \baa d x (t)=f( x (t),t)dt+b( x (t),t)dw(t). \label{Ito}\eaa By
Theorem \ref{Th2}, the process $ x $ can be approximated in $L_q$-norm by the solutions of  ordinary differential
equations (\ref{u}), where $y\in\Y$.
\end{corollary}
\par
 The approximation of solutions of It\^o stochastic differential equations by the solutions of ordinary differential
 equations with binary noise  could be a useful addition to
the existing methods of discretization such as Euler-Maruyama
discretization; see,  e.g.,  \cite{AK,Hi,K}. \index{ Abramov {\em et
al} (2011), Higham {\em et al} (2002), or Kloeden and Platen
(1992).}
\begin{remark} Theorem \ref{Th2}  and Corollary \ref{corr1}  imply that the solutions of It\^o equations (\ref{Ito})
are statistically indistinguishable from the solutions of ordinary equations (\ref{u}) with small enough $\d=t_{k+1}-t_k$, given
the presence of an arbitrarily small errors in the  measurements of the processes.
A  related feature is discussed in detail in Section 4 below.
\end{remark}

\section{Some modifications}\label{secM}

\subsection{Approximation using piecewise constant
processes}\label{secV} We assume again that $ x \in X$ is given, and
that $\F_t$ is the filtration generated by $ x $.

Let $\Z_n$ be the set of pathwise right-continuous
piecewise-constant $\F_t$-adapted real valued processes $y(t)$ such
that $y(0)= x (0)$ and that there exists $d>0$ such that  either
$y(t)=y(t_k)+d$ for $t\in[t_k,t_{k+1})$ or $y(t)=y(t_k)-d$ for
$t\in[t_k,t_{k+1})$, where $t_k=kT/n$, $k=0,...,n$, $n=1,2,...$.
The sequence $\{y(t_k)\}$ represents a path of a binomial tree
again.

Let $\Z=\cup_{n\ge 1}\Z_n$.

\begin{theorem}\label{Th3} The statement of Theorem \ref{ThM} holds with $\Y$ replaced by $\Z$.
\end{theorem}
\par
{\em Proof of Theorem \ref{Th3}} repeats the proof of Theorem \ref{ThM}
with the following changes.  For $n=1,2,...$, let $t_k=kT/n$, $k=0,...,n$, and $\d=t_{k+1}-t_k$, we define
processes
$y(t)=y_{n,m,p}(t)$  such that $y(t)=x(0)=\PP_{m,p}(0)$ for $t\in[t_0,t_1]$,
$y(t)=y(t_k)+M_{m,p}\d$ for $t\in[t_k,t_{k+1})$ if $y(t_k)\le
\PP_{m,p}(t_k)$, and $y(t)=y(t_k)-M_{m,p}\d$ for
$t\in[t_k,t_{k+1})$, if $y(t_k)\ge \PP_{m,p}(t_k)$, $k=1,2,...$. Here $M_{m,p}=2mp$.
Similarly to the proof of Theorem \ref{ThM}, we obtain that
\baaa |y(t_k)-\PP_n(t_k)|\le 2M_{m,p}\d, \quad k=0,1,...,n.
\label{ineq3}\eaaa
It follows that \baaa |y(t)-\PP_n(t)|\le 4M_{m,p}\d, \quad t\in[0,T].
\label{ineq4}\eaaa
The remaining proof repeats the proof of Theorem \ref{ThM}. $\Box$
\vspace{3mm}
\par
Further, let
 $f(x,t):\R\times[0,T]\to\R$ be a continuous function
 that is bounded together with the derivative  $\p f(x,t)/\p x$.
Let  $\V_n$ be the set of real
valued processes $v(t)$ such that $u(0)= x (0)$ and that \baa
v(t)=v(0)+\int_0^tf(v(s),s)ds+z(t), \label{v}\eaa where $z\in\Z_n$.

Clearly, $v$ is uniquely defined; it is a $\F_t$-adapted  process with jumps at the times $t_k$.

Let $\V=\cup_{n\ge 1}\V_n$.

The following theorem represents a modification  of Theorem \ref{Th2}.
\begin{theorem} \label{Th4} The statements of Theorem \ref{Th2} and Corollary \ref{corr1} hold with $\U$ replaced by $\V$.
\end{theorem}
\par
\par
Again, Theorem \ref{Th4} applied with $f\equiv 0$  gives
Theorem \ref{Th3}; therefore, Theorem \ref{Th4} represents a  generalization of Theorem
\ref{Th3}.
\par
{\em Proof of Theorem \ref{Th4}} repeats the proof of Theorem
\ref{Th2} with the changes similar to the changes that were done in
the proof of Theorem \ref{Th3}. $\Box$
\subsection{Binomial approximation of $\log x(t)$}\label{secLog}
In financial modelling, it is common to approximate positively valued stochastic processes by binomial processes with the rate of change decreasing near zero such that their logarithm have the constant rate of change. We need to modify our approach to cover these problems.

Let $ x \in X$ be given such that $x(t)>0$ for all $t$ and that the process
$\log x(t)$ belongs to  $X$. Let $\F_t$ be the
filtration generated by $ x (t)$.

Let $\Y^+_n=\Y^+_n(x(\cdot))$ be the set of pathwise continuous piecewise
affine  $\F_t$-adapted real valued processes
$y(t)$ such
that $y(0)= x (0)$ and that there exists $d_1\in(0,\d)$ and $d_2>0$  such that   either $y(t_{k+1})=y(t_k)(1-d_1\d)$
or $y(t_{k+1})=y(t_k)(1+d_2\d)$. Here $t_k=kT/n$,
$k=0,...,n$, $n=1,2,...$, and $\d=t_{k+1}-t_k=T/n$.

Let $\Y^+=\cup_{n\ge 1}\Y^+_n$, and let $\Y^+_{\logs}$ be the set of all processes $\eta(t)$ such that $\eta(t)=\log y(t)$, where $y\in\Y^+$.

In financial modelling, binomial processes
from $\Y^+$ are used for positively valued
stochastic It\^o processes with lognormal distributions describing
the evolution of the stock prices in the Black-Scholes market model.
We will be using these processes in the next section addressing the financial applications.

\begin{theorem}\label{ThLog} The statement of
 Theorems \ref{Th3}  holds with  $ x (t)$ replaced by $\log x(t)$ and  with $\Y$ replaced by $\Y^+_{\logs}$ .
\end{theorem}

{\em Proof.} The proof requires a small modification of the proof
of  Theorems \ref{Th3}. We select $d_1$ and
$d_2$ such that $\log(1-d_1\d)=-M_{m,p}\d$ and
$\log(1+d_2\d)=M_{m,p}\d$, where $\d=t_{k+1}-t_k$ and $M_{m,p}=2mp$ are
selected similarly to the proof of Theorem \ref{ThM}.
We define $\eta\in\Y^+_{\logs}$ by selecting
\baaa
\eta(t_k)=\log x(t_0)+\sum_{i=1}^k\xi_i\d,\eaaa where $\xi_i$ take values $\pm M_{m,p}$ selected similarly to the proof
of  Theorems \ref{ThM}. Consider representation
$\xi_i\d=\log(1+\zeta_i\d)$, where $\zeta_i$ take values $-d_1$ or $d_2$.
We have that
\baaa
\eta(t_k)=\log x(t_0)+\sum_{i=1}^k\log(1+\zeta_i\d).\eaaa
Then $\eta(t)=\log y(t)$, where $y\in\Y^+$ is such that
 $y(t_0)=x(t_0)$ and
$y(t_k)=x(t_0)\prod_{i=0}^k(1+\zeta_i\d)$ for $k>0$. This completes the proof.
   $\Box$

 A similar result can be obtained for piecewise constant approximations.
\subsection{Approximation with dynamically
adjusted sizes of the binary increments}\label{secVar}
  It could be interesting to consider approximating sequences of processes  with dynamically adjusted
sizes of the increments  that can replicate
the changes in the evolution law  for the underlying process, such as  the dynamics of the volatility for the stock prices.
So far, Example \ref{exa1}(ii) shows that this feature is not feasible for the approximating processes from $\Y$.

It appears that, for the case of  underlying processes from some more narrow  classes, the algorithm described in Theorem \ref{ThM}
 can be extended on the approximating processes with dynamically adjusted
sizes of the increments.

Let $H_{\t,q}$ be the set of all processes $ x \in X$ such that  there exists $\e_0>0$,  $q\in (0,1]$, $\t>0$, $C>0$, and a $\F_t$-adapted stochastic process  $\s(t)$  such that $0\le \s(t,\o)\le C$
for all $t\in[0,T]$ and $\o\in\O$,
and that
\baa\sup_{(t,\o)\in[0,T]\times\O}\frac{| x (t,\o)- x (t-\e,\o)|}{\e^q}\le \s(t-\t,\o)
\label{Hol}
\eaa
for all $\e\in(0,\e_0)$.

For $ x \in H_{\t,q}$, the approximating processes  with dynamically adjusted sizes of binary increments can be constructed as follows.
For $n=1,2,...$, we select  $y(t)=y(t_k)\pm
\d^{q-1}\s(t_k,\o)(t-t_k)$, $t\in[t_k,t_{k+1})$, where
$\d=t_k-t_{k+1}=T/n$.  We can skip construction of  the
processes $\QQ_m$.  We construct the process  $\PP_{m,p}(t)$ using $ x $
instead of $\QQ_m$, and observe that $\esssup_t|d\PP_{m,p}(t)/dt|\le
\e^{q-1}\s(t-\t)$ for $\e=1/p$. The approximating  properties can be
established  as before, with sufficiently small  $\d\le
\min(\e_0,\t)$ .

The H\"older type condition (\ref{Hol}) for  $H_{\t,q}$ is close to
the H\"older regularity property for the trajectories of the
continuous It\^o processes; see, e.g., \cite{I73}. \index{Ibragimov
(1973).} Unfortunately,   the presence of the supremum over $\o$ still
makes condition (\ref{Hol})  too restrictive; this condition   is
not satisfied even for It\^o processes with constant diffusion
coefficients, including a Wiener process.

Condition (\ref{Hol}) could be reasonable for  models   using
a causal estimator for $\s$ based on
historical observations of $ x $, under  a hypothesis that the
currently calculated  $\s(t,\o)$ satisfies  (\ref{Hol}) on the time
interval $[t,t+\t]$.  Then condition (\ref{Hol}) ensures that
\baaa\sup_{(t,\o)\in[0,T]\times\O}\frac{| x (t+\tau,\o)- x (t+\tau-\e,\o)|}{\e^q}\le
\s(t+\tau-\t,\o) \label{Hol2} \eaaa for all $\e\in(0,\e_0)$ and
$\tau\in[0,\t]$. Since $\s(t+\tau-\t)$ is a $\F_t$-measurable random
variable, we have that  the value of the upper boundary is known at
time $t$.

  If we  interpret $\s(t-\t)$
as an analog of the diffusion coefficient at time $t$, then
condition (\ref{Hol}) can be interpreted as the requirement that
the diffusion coefficient is predictable on the time horizon $\t$.
In other word, this is a requirement that there is some stability in
the evolution law for $ x $. This could be a reasonable requirement
for many models.

 \section{Applications for the  financial modelling}\label{secF}
In quantitative finance, the classical discrete time
Cox-Ross-Rubinstein  model of a single-stock  financial market includes a bond or
money market account with the price $B_k$ and a single risky asset
with the price $S_k$. In this model,  the process $B_k$ is assumed to be non-random
or risk-free and is used as a num\'eraire, and $S_k$ is assumed to
be a binomial stochastic process, $k=0,1,2,...$.  For simplicity, we assume that $B_k=\rho^k$ for some $\rho\ge 1$. Let $\ww S_k=B_k^{-1}S_k$ be the discounted price process.  For the
Cox-Ross-Rubinstein  model, $\ww S_{k+1}=\ww S_k(1+\zeta_{k+1})$, where
$\zeta_k$ takes only two values, $-d_1$ and $d_2$, such that $d_1\in (0,1)$ and $d_2>0$; see, e.g., \cite{Dbook}, Chapter 3, and \cite{D07}.   This  model is a
so-called complete market where any claim can be replicated and
where there is a unique martingale (risk-neutral)  measure
equivalent to the historical measure. For complete market models,
the price of a derivative is defined via the expectation of the
payoff by this unique martingale measure.

\def\S{\textsf{S}}
\def\B{{\rm B}}\def\HH{{\bf H}}
We consider also a  continuous time model that includes a bond or money market
account with the price $\B(t)=e^{rt}$ and a single risky asset with the
price $\S(t)$, $t\in[0,T]$. Here $r\ge 0$ is given and known, and $\S(t)$ is
assumed to be a stochastic Ito process such that
\baaa
d\S(t) =\S(t)[a(t)dt+\s(t) dw(t)],
\eaaa
where $a(t)$ is some appreciation rate process, $\s(t)=\s(t,\o)$ is a random volatility process, and $w(t)$ is a Wiener process.  We assume that $\s(t)$ is  independent on
the increments $w(\t)-w(\tau)$, $\t>\tau\ge t$.
We assume for simplicity that $a\in\R$ is a constant.

Let $\ww \S(t)=\B(t)^{-1}\S(t)$ be the discounted price process.

The classical Black and
Scholes continuous time model  represents a special case of this model  with a constant volatility $\s>0$. In this case, the market model is complete; see, e.g.,  \cite{Dbook}, Chapter 5. The
Donsker theorem allows to   approximate the process $\ww \S(t)$ by binomial processes $y\in \Y^+$ in distributions.
This allows to replace pricing of derivatives in the continuous time setting by the pricing in the discrete time setting via binomial trees (i.e., finite differences).

The pricing of derivatives is usually more difficult for the so-called incomplete market models where
a martingale measure is not unique. Some important examples of
market incompleteness arise when the volatility $\s=\s(t,\o)$  is time varying, random, and not adapted to the filtration generated by $w(t)$.  In this case, a straightforward discrete time
approximation leads to  discrete time market models
   such as binomial models  with dynamically
adjusted sizes (i.e., random sizes) of the binary increments; see,
e.g., \cite{AD} \index{  Akyildirim {\em et al}  (2012),} and
discussion in Section \ref{secVar}. These binomial  models are
incomplete.

On the other hand,    Theorem \ref{ThLog} implies  that, for any continuous time market model, including models with random
    volatility $\s=\s(t,\o)$, there exists a process  $y\in\Y^+$
     such that $y$ and $\ww\S$  are statistically indistinguishable, given
 that there is a non-zero measurements error, for instance, a rounding error, or any arbitrarily small error.

    This $y$ can be used for construction of a complete discrete time market that approximates the original incomplete market as follows.

For $n=1,2,...$, consider the sets $\{t_k\}_{k=0}^n$
such that $t_k=k\d$, $\d=T/n$.

Let $B_k=e^{rt_k}=\B(t_k)$,
$\ww S_k=y(t_k)$, and $S_k=B_k\ww\S(t_k)$.
Consider a discrete time market model with the stock prices $S_k$, with the discounted prices $\ww S_k$, and with the bond prices $B_k$. By the definitions, this is a Cox-Ross-Rubinstein model, i.e., a complete market model.

This leads to a counterintuitive conclusion that the incomplete
markets are indistinguishable from the complete markets by
econometric methods, i.e., in the terms of the market statistics.

Let us elaborate this conclusion.
    It is  known that the market completeness is
not a robust   property: small random deviations of the coefficients
convert a complete market model into a incomplete one. Thanks to
Theorem \ref{ThLog} and approximation scheme described above, we can claim  now  that market incompleteness is
also non-robust: small deviations can convert an incomplete model
into a complete one. More precisely, it implies that, for any
incomplete market from a wide class of models, there exists a
complete market model with arbitrarily close discrete sets of the observed
processes.

It can be further illustrated  as the following.   Assume that we
collect the marked data (the prices) for
$t\in[0,T]$, with the purpose to test the following
hypotheses $\HH_0$ and $\HH_A$ about the stock price evolution:
\begin{itemize}
\item[$\HH_0$:]  For any sampling interval $\d$, the discrete time market
with the stock prices $S_k=\S(t_k)$ is incomplete; and
\item[$\HH_A$:]
There exists  a sampling interval $\d$ such that the discrete time market
with the stock prices $S_k=\S(t_k)$ is complete.
\end{itemize}
Here  $\d=T/n$, $n=1,2,...$, $t_k=k\d$, $k=0,1,...,n$.

According to Theorem \ref{ThLog},  it is impossible to reject
$\HH_A$ hypothesis based solely on the market data collected, for a random volatility process $\s(t,\o)$
generating an incomplete continuous time market and incomplete discrete time markets based on the sampled prices.

\par
It can be noted that we can replace the hypothesis $\HH_0$ by a
hypothesis assuming a particular stochastic price model, such as a
Markov chain model for the volatility,  Heston  model, etc.

Due to rounding errors, the statistical indistinguishability leading
to this conclusion cannot be fixed via the sample increasing since
the statistics for the incomplete market models can be arbitrarily
close to the statistics of  the alternative complete models.

It must be clarified that this conclusion has rather a purely
theoretical value. Unfortunately, the causal binomial approximation described above
 is not particularly useful for practical options pricing since the
process $\{S_k\}$ does not represent the price of a tradable assets in
the market model with the price $\{\S(t_k)\}$; the values $S_k$
represent  the prices of a tradable assets    for the new discrete
time market model only. In addition,  the hedging strategies for the discrete time market with the prices  $S_k$
do not give the same output when applied to the  prices samples of the original prices $\S(t_k)$, since  small errors for
single transactions could be accumulated into a significant error even  if $S_k$ is  close to $\S(t_k)$ a.e..
This is because a close approximation requires small sampling intervals, large number of periods,
and  a large number of transactions. Each  transaction will  generate  a small but non-zero error
since we allow that $S_k\neq \S(t_k)$ even if  these values are close.
Furtehrmore,  using of the new discrete time complete market model with the price process
$\{S_k\}$  for the options
pricing would lead to overpricing if $\s(t,\o)$
is  random and time variable, since the process $S_k$ is constructed such
that its rate of change  is constant; see Example \ref{exa1}. It can be noted that  approximation in the
class $H_{\t,q}$ described in Section \ref{secVar} allows to
replicate stochastic and time varying volatility; however, this approximation does
not lead to approximating complete markets.
\par
The results of this section on statistical indistinguishability of the complete binomial markets and incomplete markets was presented  on The Quantitative Methods in Finance conference in Sydney in December
2013.  A related result was obtained in \cite{D12}, where the approximation  was considered in a diffusion setting
 that allowed to approximate  the dynamics of the original volatility as well, via approximation by diffusion processes.
 \section{Discussion and  future development}
The  approach suggested in this paper  allows many more
modifications. We outline below some possible straightforward
modifications as well as
  more challenging problems and possible applications  that we leave for the future research.
\begin{enumerate}
\item
Instead of  binomial processes, other processes could be used  for approximation, for
instance, trinomial processes.
\item It  could be interesting  to  extend the construction from Section \ref{secVar} on a case where  condition (\ref{Hol}) is replaced by a weaker condition that covers It\^o processes.
\item Ordinary differential equations (\ref{u}) and  equations with jumps (\ref{v}) can be investigated
with random noise with preselected distributions rather than with the processes  $y(t)$ and $z(t)$ defined by
the approximation procedure.   For example,  equation (\ref{u})  could be considered  for a binary white  noise
$\eta=dy/dt$ that is a piecewise constant stochastic
process.  Equation (\ref{v})  could be considered  for a random input represented by binary
white noise $\eta=dz/dt$ such that $\eta(t)=\pm c\,${\Large $\delta$}$(t-t_k)$ for
$t\in (t_{k-1},t_{k+1})$, where
 {\Large $\delta$}$(t)$ is a delta-function,
$c=c_\eta\in\R$.
\item
For  equations (\ref{u}) and (\ref{v}) with random noise, optimal stochastic control problems as well as  stability and instability could be investigated,
as was done for  Euler-Maruyama discretization; see, e.g.,
\cite{RD} and the bibliography there.
\item
Approximation  of classical stochastic differential plant equations  by equations (\ref{u}) or (\ref{v})   with random noise
could be applied  for solution of optimal stochastic control problems,   in particular, for optimal portfolio selection
problems. Some related results for  Euler-Maruyama discretization were obtained in
\cite{RDD}.
\item It could be useful to investigate the binomial approximation in relation to the  transmission of digital signals such as described in \cite{DS},\cite{DS2}.
\end{enumerate}

\section*{Acknowledgment} This work  was supported by Australian Research Council grant
 DP120100928 to the author.

\end{document}